\documentclass{vospwks2007}
\usepackage{graphicx}

\title{How to use the SEDs produced by synthesis models (inside and outside the VO)?}
\author[1,2]{Miguel Cervi\~no}
\author[1,2]{Valentina Luridiana}
\affil[1]{Spanish Virtual Observatory (SVO; Spain)}
\affil[2]{Instituto de Astrof\'\i sica de Andaluc\'\i a (IAA-CSIC; Spain)}

\begin{document}


\maketitle

\begin{abstract}
In this contribution we investigate how to describe the results and usage of evolutionary synthesis models. In particular, we look for an explicit and quantitative description of the parameter space of synthesis models and the evaluation of their associated uncertainties and dispersion. First, we need to understand what synthesis models actually compute: we show that a synthetic stellar population with fixed physical parameters (age, metallicity, star formation history, initial mass function and {\it size} of the system) can only be described in terms of probability distributions (i.e. there is an intrinsic dispersion in any model). Second, we need to identify and characterize the coverage in the parameter space of the models (i.e. the combinations of input parameters that yield meaningful models) and the different sources of systematic errors. Third, we need a way to describe quantitatively the intrinsic dispersion, the systematic error and the parameter space coverage of the models.

Up to now, the parameter space coverage and uncertainties have been described qualitatively in the models' reference papers, with potential misinterpretations by models' users. We show how Virtual Observatory developments enable a correct use of synthesis models to obtain accurate (and not simply precise) results. 

\end{abstract}

\section{Introduction}

The aim of the Virtual Observatory (VO) is to provide an environment to produce more accurate science. Not only does the VO provide an opportunity for technical developments, it also offers unique possibilities for scientific research: The VO framework requires an explicit description of the data that it provides with no implicit assumptions about the data themselves. This requirement is fundamental since interoperability is performed by machine to machine communications without supervision by human beings. Such a description is performed by the data model working group in the International Virtual Observatory Alliance (IVOA).

The rationale of the data model working group is to provide an abstract description of concepts and their interrelationships, used to fix both the names and meanings of concepts in the VO context and also their internal structure and cross-connections (IVOA Spectral Data Model 1.01, 2007)\footnote{\tiny{\tt http://www.ivoa.net/Documents/WD/DM/SpectrumDM-20070515.html}}, that is to describe all the possible universes of classes of data. Note that such description and interrelationships are fundamental in scientific research also, with the only difference that, in the last case, researchers are focused on {\it particular} types of objects.

\begin{figure}
\centering
\includegraphics[width=0.8\linewidth]{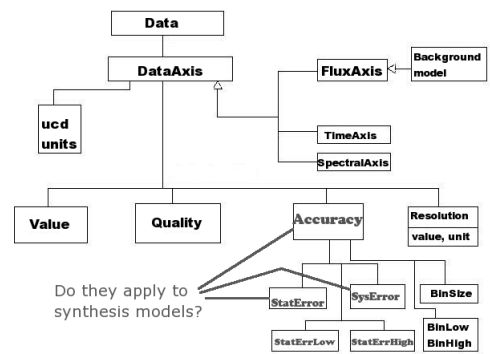}
\caption{IVOA spectrum data model.} 
\label{fig:IVOA-SDM}
\end{figure}

Figure \ref{fig:IVOA-SDM} shows the spectrum data model generated by the data model working group. This data model aims to describe all the universe of tables in the VO that contain a spectrum, hence the data model must not only be valid for observed data, but also for theoretical spectra. 

This data model (the generic description of a spectrum) allows to formulate new questions for researchers that either produce or use theoretical spectra: Does this data model describe theoretical spectra? Can all the ``boxes'' in the data model be filled? And, if this is the case, what is the accuracy of a theoretical model? Do theoretical models have systematic errors? And, do they have statistical errors? 

\begin{figure}
\centering
\includegraphics[width=0.8\linewidth]{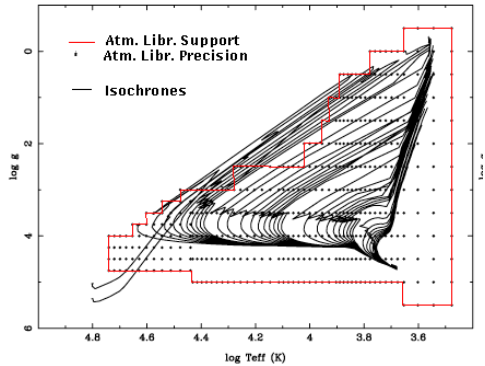}
\caption{Draft illustration of the concepts of the IVOA characterization data model for a typical atmosphere library like those used in populations synthesis showing the resolution of the library ({\it precision} in the VO data model) and its coverage ({\it support} in the VO data model). Isochrones are also drawn for an intuitive comparison of the support of both sets of data.} 
\label{fig:IVOA-CDM}
\end{figure}

In addition to the spectrum data model, this working group has also proposed a data model for astrophysical dataset characterization (IVOA Data Model for Astronomical DataSet Characterization 1.11, 2007)\footnote{\tiny{\tt http://www.ivoa.net/Documents/PR/DM/CharacterisationDM-20070530.html}}. This data model defines the information required to describe the parameter space of observed or simulated astronomical datasets. Fig. \ref{fig:IVOA-CDM} is a draft illustration of the description of an atmosphere library like those used in population synthesis.

Again, a correct characterization of models is fundamental for their application to observations. All theoretical models have a limited range of validity, given by the computational algorithms, the resolution of the input data and other factors. Usually, these limitations are described qualitatively in the reference papers, but unfortunately, it is not so common for these limitations to be introduced in the code that computes the models \cite[Fortunately, there are some significant exceptions, like CLOUDY by][]{cloudy}. 

We stress again that an answer of the previous issues is fundamental in scientific research, since the basic way to obtain the physical parameters of an observed system is to compare observational data with theoretical models. Only if these questions are solved in a quantitative way, will it be possible to evaluate the accuracy (not the precision) of our knowledge of the observed system. 

In the case of stellar populations, theoretical models are provided by stellar population synthesis \citep{fanelli,tinsley}. The general case of population synthesis aims to decompose the observed light of the system under study in different stellar populations components, that is the contribution of classes of stars, each class with different effective temperature and gravity \citep{fanelli}. In this case, no further information about the observed system can be obtained unless additional a priori information is considered. In the case of {\it evolutionary} population synthesis \citep{tinsley}, the contribution of the different stellar classes is defined by the evolutionary status of the system, hence the knowledge of the contribution of the different stellar classes can be transformed into knowledge about the age, metallicity, stellar birth-rate (i.e. star formation history, initial mass function and the amount of formed stars) of the system. Although evolutionary synthesis models are powerful, it is necessary to take into account that:
\begin{itemize}
\item The transformation between the contribution of stellar classes and the associated physical parameters requires some a priori hypotheses: a set of atmosphere models, evolutionary tracks and defined star formation properties (i.e. stellar birth rate).
\item There are different evolutionary conditions that result in the same integrated spectra, so the physical parameters obtained from the analysis of observations are not necessarily unique. 
\end{itemize}

These a priori hypotheses, together with the computational algorithms and approximations used by different synthesis codes, define the range of use (i.e. characterize) of the results. They are also the principal source of systematic errors in model computations. On the other hand, the degeneracy of the possible solutions produces an intrinsic precision loss in any fit of observational data and theoretical models: the best-fit solutions are not necessarily the correct ones and the distribution of the goodness of fit is also needed for a correct interpretation of the observed data. In order to analyze these issues in detail, we need to understand what a synthesis model is.

\section{What is a synthesis model?}

The first step needed to evaluate systematic errors, dispersion and the range of application in population synthesis models is to understand how synthesis models work. The main result of synthesis models is the luminosity due to an ensemble of $n$ individual stars, each with a given luminosity, $\ell_{i}$. The total luminosity of the ensemble is:

\begin{equation}
L_{\mathrm{tot}} = \sum_{i=1}^n \ell_{i} = n\, \frac{1}{n} \sum_{i=1}^n \ell_{i} = n\, <\hat{\ell}>.
\label{eq:Ltot}
\end{equation}

\noindent where $<\hat{\ell}>$ is the {\it estimation of the mean} luminosity of an individual star.

However, in a system where individual stars are not resolved, we cannot perform such a sum, since we do not know the individual $\ell_{i}$ values. In this case, we can obtain the total luminosity as the composition of different stellar populations (or different stellar types). If we assume that a given stellar type $i$ {\it can be represented} by a luminosity $l_{i}$ and we have some rules that give us the proportion $w_{i}$ between different stellar types for different physical parameters ($w_{i}=w_{i}[t, \mathrm{Z}, \mathrm{IMF}, \mathrm{SFH}]$), then

\begin{eqnarray}
L_{\mathrm{tot}}^{\mathrm{theo}}(n) &=& n\, \sum_{i=1}^{N_{\mathrm{class}}} w_{i}\, l_{i} = n\,  <\ell>,\label{eq:LSSP}
\\
&&\sum_{i=1}^{N_{\mathrm{class}}} w_{i} = 1.\label{eq:LSSPnorm}
\end{eqnarray}

\noindent where $<\ell>$ is the {\it mean} (or average) luminosity of an average star that represents the ensemble. Note that this mean luminosity is not necessarily the luminosity of any {\it real} star, but it is a description of an underlying probability distribution of the possibles luminosities of real stars in the ensemble. As a trivial example, remember the half-alive half-dead Schr\"odinger's cat solution in quantum mechanics.

Although Eq. \ref{eq:Ltot} and Eq. \ref{eq:LSSP} look similar, they do not provide the same quantity: The first case is the {\it actual} integrated luminosity of a given cluster and the second case is the {\it mean} integrated luminosity of all the possible clusters with $n$ stars and in the same evolutionary conditions. Note also that the problem in population synthesis is not the classical statistical one of estimating $n\,  <\ell>$, but the inverse one: obtaining the {\it possible} theoretical $n\,  <\ell>$ values compatible with the observed integrated luminosity $L_{\mathrm{tot}}$. So, it is necessary to use the complete distribution of theoretical integrated luminosities rather than its mean value $L_{\mathrm{tot}}^{\mathrm{theo}}(n)$. This distribution provides a {\it metric for goodness-of-fit} to stellar population studies.

\begin{figure}
\centering
\includegraphics[width=0.8\linewidth]{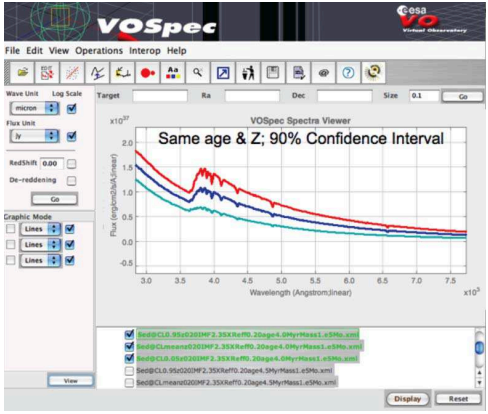}
\caption{Mean spectrum and 90\% confidence interval region of clusters with $10^5 M_{\odot}$ transformed into stars, with a given metallicity and a given age.} 
\label{fig:SSPmean}
\end{figure}

\begin{figure}
\centering
\includegraphics[width=0.8\linewidth]{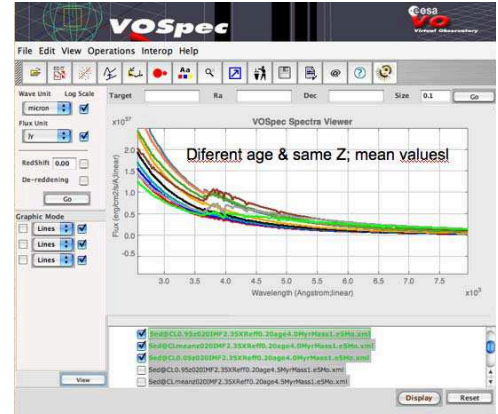}
\caption{Mean spectrum of clusters with $10^5 M_{\odot}$ transformed into stars, with a given metallicity and a wide range of ages.} 
\label{fig:SSPvariance}
\end{figure}

As an illustration of the effects of this intrinsic dispersion, we show in Fig. \ref{fig:SSPmean} the 90\% confidence interval of the integrated luminosity of a stellar cluster with $10^5 M_{\odot}$ transformed into stars for a given stellar birth rate, metallicity and age. It represents the integrated spectrum of {\it all} possible clusters with this number of stars, and evolutionary conditions. Fig. \ref{fig:SSPvariance} show the mean spectrum of a cluster with similar conditions but different ages. The 90\% confidence interval in Fig. \ref{fig:SSPmean}, with a single age, covers all the region defined by the mean values of the spectra at different ages. Any fit of a real spectrum of such a cluster would produce a very precise result for the age, but the accuracy of this fit cannot be known if the underlying probability distributions of integrated luminosities at different ages are not taken into account. 

We refer to \cite{CLCL} for a more extensive analysis of the probabilistic formulation of population synthesis and \cite{vallarta} for a more complete description about the metrics for goodness of fit.
  
\section{Systematic errors in synthesis models}

Once the origin of the intrinsic dispersion of population synthesis models has been described, we are going to discuss the sources of systematic errors and how to take advantage of the IVOA Characterization data model to define the region of application of synthesis models in its parameter space. We refer to \cite{cancun} for a more complete review on the systematic errors.

In the following, we will only consider the systematic errors related to stellar population synthesis computations itself, assuming that the systematic errors of the inputs are known. The main source of systematic errors are:

\begin{enumerate}
\item Atmosphere libraries $l_{i}$: systematic errors in atmosphere libraries mainly depend on how realistic our assumption about the classification of stars in the different stellar types are (see Garc\'\i a-Vargas' et al. contribution in these proceedings). Such an evaluation must be implemented by synthesis models makers.

In addition to the systematic errors, the range of application of synthesis models depends in the coverage on the parameter space ($\log g$, $T_{\mathrm{eff}}$, metallicity, ...), that is the characterization of the library in this parameter space.

\item Tracks/Isochrones and their combination with the star formation history $n_{i}$: In the case of synthesis models, the shape of isochrones and the {\it density} of stars along the isochrone must fit observational data.  Small variations in these densities can produce big differences in the models. 
Hence, it is crucial to know the uncertainties in the evolutionary tracks/isochrones used to evaluate the systematic error in synthesis models. Currently, there are different groups working on the evaluation of systematic errors in evolutionary tracks, like Degl'Innocenti et al. and Bressan et al. 

Additionally, since synthesis models describe the whole population, they also need to take into account evolutionary phases that are included in standard evolutionary tracks, like pre-main sequence evolution, final evolutionary stages like SN or cooling white dwarfs, etc. In general, synthesis models makers combine tracks of different groups to obtain a good coverage in evolutionary phases, but in doing so, they include a potential source of error due to different physics used in track computations. The evaluation of the associated systematic errors and the coverage of the models results due to these effects must be implemented by synthesis models makers.

\item Synthesis models algorithms: the final issue is the algorithms used in the model. The main problem is how to combine the different coverage and sampling (support and precision respectively, following the Characterization data model) of the atmosphere libraries and isochrones in the $\log g -\log T_{\mathrm{eff}}$ plane (see Fig. \ref{fig:IVOA-CDM}). In some cases, it is necessary to extrapolate the behavior of atmosphere libraries to regions not covered by the libraries. In addition, there are also differences in how the correspondence of the points in the isochrones and in the atmosphere libraries is done, ranging from how the libraries are interpolated to how the closest model to the given isochrone point is chosen.

\end{enumerate}

Although the characterization of synthesis models is a task under development, it is currently possible to define its range of application by comparing the coverage of isochrones and atmosphere libraries and using the IVOA characterization data model for this task. It will not be the final solution of the problem of coverage, but it will enormously help towards the use of synthesis models in their correct range of application.

\section{Conclusions}

Synthesis models, as any other theoretical model, are affected by systematic errors. Although the evaluation of such errors is a difficult task that only can be performed by synthesis models makers, the current IVOA description of spectrum data model allows to add these errors to the output.

In the case of statistical errors, the spectral data model does not apply directly to theoretical results since the goal of statistics is inferring the underlying distribution from which the data have been drawn, which is assumed to be known in the computation of theoretical models. However, the {\it StatError} box in the current spectral data model can be used to describe the {\it intrinsic dispersion} of theoretical models. The dispersion of synthesis models, although not recognized as such, is computed by several codes under the label of surface brightness fluctuations, so this piece of information is already available.

Finally, a proper characterization of isochrones and atmosphere libraries coverage allows to obtain a realistic characterization of synthesis models, and its range of application.

\section*{Acknowledgments}

This research has made use of the Spanish Virtual Observatory supported by the Spanish MCyT through grants AyA2005-04286 and AyA2005-24102-E. It has also been supported by the Spanish MCyT through the project AYA2004-02703. MC is supported by a {\it Ram\'on y Cajal} fellowship.

\end{document}